\documentclass{article}
\usepackage{dcase2019,amsmath,graphicx,url,times,booktabs, tabularx,epstopdf,multirow}

\title{Acoustic Scene Classification Based on a Large-margin Factorized CNN}

 \name{Janghoon Cho,
 Sungrack Yun,
 Hyoungwoo Park,
 Jungyun Eum,
 Kyuwoong Hwang}

 \address{Qualcomm AI Research\sthanks{Qualcomm AI Research is an initiative of Qualcomm Technologies, Inc.}, Qualcomm Korea YH\\
 343, Hakdong-ro, Gangnam-gu, Seoul, Korea\\
 \{janghoon, sungrack, c\_hyoupa, c\_jeum, kyuwoong\}@qti.qualcomm.com\\
}

\begin{document}

\ninept
\maketitle

\begin{sloppy}

\begin{abstract}
In this paper, we present an acoustic scene classification framework based on a large-margin factorized convolutional neural network (CNN). We adopt the factorized CNN to learn the patterns in the time-frequency domain by factorizing the 2D kernel into two separate 1D kernels. The factorized kernel leads to learn the main component of two patterns: the long-term ambient and short-term event sounds which are the key patterns of the audio scene classification. In training our model, we consider the loss function based on the triplet sampling such that the same audio scene samples from different environments are minimized, and simultaneously the different audio scene samples are maximized. With this loss function, the samples from the same audio scene are clustered independently of the environment, and thus we can get the classifier with better generalization ability in an unseen environment. We evaluated our audio scene classification framework using the dataset of the DCASE challenge 2019 task1A. Experimental results show that the proposed algorithm improves the performance of the baseline network and reduces the number of parameters to one third. Furthermore, the performance gain is higher on unseen data, and it shows that the proposed algorithm has better generalization ability.
	
\end{abstract}

\begin{keywords}
Acoustic scene classification, Factorized convolutional neural network, Triplet sampling
\end{keywords}

\section{Introduction}
\label{sec:intro}
The interest of acoustic scene classification (ASC) has been continuously increasing in the last few years and is becoming an important research in the fields of acoustic signal processing. 
The ASC aims to identify different environments given the sounds they produce \cite{barchiesi2015acoustic} and has various applications in context-awareness and surveillance \cite{valenti2016dcase,radhakrishnan2005audio,chu2009environmental}: e.g. the device which recognizes the environmental sound by analyzing the surrounded audio information.
With the release of large scale datasets and challenge tasks by Detection and
Classification of Acoustic Scenes and Events (DCASE) \cite{mesaros2018multi, DCASE2019TASK1A}, the ASC has become a very popular research topic in audio signal processing. We have an increasing number of research centers, companies, and universities participating in the DCASE challenge and workshop every year. 

In the past decade, deep learning has accomplished many achievements in audio, image, and natural language processing. Especially, the algorithms based on the convolutional neural network (CNN) are dominant in ASC \cite{phaye2019subspectralnet,barchiesi2015acoustic,han2017convolutional,salamon2017deep} tasks. 
In \cite{valenti2016dcase}, a simple CNN with 2 layers was adopted, and many attempts were tried to solve the overfitting problem in improving the ASC performance by increasing the model complexity: e.g. number of layers in CNN. In \cite{phaye2019subspectralnet}, SubSpectralNet method was introduced to capture more enhanced features with a convolutional layer by splitting the time-frequency features into sub-spectrograms. In \cite{han2017convolutional}, a simple pre-processing method was adopted to emphasize the different aspects of the acoustic scene. In \cite{sakashita2018acoustic, dorfer2018acoustic, zeinali2018convolutional}, an ensemble of various acoustic features such as MFCC, HPSS, i-vector and the technique that independently learns the classifiers of each feature was proposed. However, the method is heuristic and requires a lot of computation in the front-end step before CNN inference step.

In order to solve the problems of the above-mentioned conventional methods, we propose an algorithm that uses only one feature and does not increase the model complexity of CNN. The pattern of each acoustic scene in the time-frequency domain can be represented as a low-rank matrix, thus we consider designing a 2D convolution layer as two consecutive convolution layers with 1D kernels. Also, we consider a loss function such that the samples from the same audio scene are clustered independently of the environment to be robust to unseen environment. In short, our proposed ASC framework is based on the Sub-Spectral Net \cite{phaye2019subspectralnet}, and the kernel factorization and loss function are applied to reduce the computational complexity and increase the generalization ability on unseen environment. We evaluated our ASC framework using the dataset of DCASE task1A, and all experimental results show that the proposed algorithm with the data augmentation techniques \cite{zhang2017mixup,park2019specaugment} significantly improves the accuracy.

The rest of the paper is organized as follows. Section \ref{sec:ASC} formulates the problem of ASC. Section \ref{sec:proposed_algorithm} describes the proposed algorithm including our factorized CNN structure and novel loss function. Section \ref{sec:experiment} presents the experimental results and analysis. Finally, Section \ref{sec:conclusion} concludes. 

\begin{figure}[t]
	\centering
	\centerline{\includegraphics[width=\columnwidth]{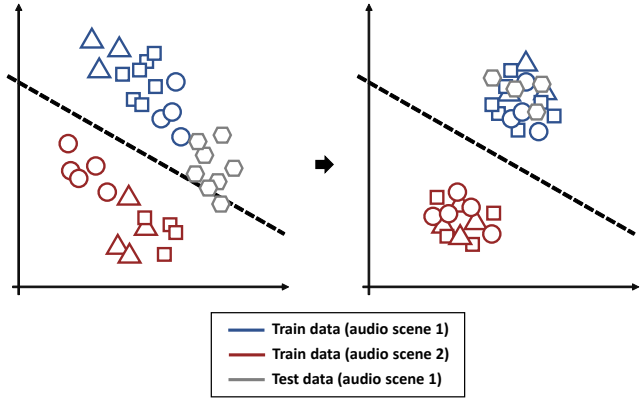}}
	\caption{Examples of two audio scene samples (red, blue) from different cities (Triangle, rectangle, circle, hexagon). Given an audio scene, the audio scene from the same city are more clustered than from different cities. For the better generalization ability on unseen city, we apply a loss such that the samples are clustered independently of the cities. }
	\label{fig:cluster_example}
\end{figure}

\section{Acoustic Scene Classification}
\label{sec:ASC}


Audio scene classification, the task1A of the DCASE 2019 challenge \cite{DCASE2019TASK1A}, is a process of predicting a label $y^*$ given an input audio clip $\bf x$ as:
\begin{eqnarray}
y^* = \arg\max_{y \in \mathcal{Y}} p(y | {\bf f}_{\bf x} ; \theta )
\end{eqnarray}
where $p(y | {\bf f}_{\bf x} ; \theta )$ is the audio scene posterior given the feature map ${\bf f}_{\bf x}$ with the network parameter $\theta$, and $\mathcal{Y}$ is the entire set of scene labels. The input audio clip $\bf x$ contains only one audio scene, and the feature map of $\bf x$, ${\bf f}_{\bf x}$, can be obtained using various algorithms such as deep audio embeddings \cite{cramer2019look, arora2019deep}, log-mel \cite{taal2011algorithm, hoshen2015speech, sakashita2018acoustic, zeinali2018convolutional}, amplitude modulation filter bank \cite{dau1997modeling, moritz2015auditory}, and  perceptual weighted power spectrogram \cite{dorfer2018acoustic}. In this paper, we use the 40 log-mel, ${\bf f}_{\bf x}\in {\rm R}^a$, since recently many approaches \cite{sakashita2018acoustic, zeinali2018convolutional} adopt the log-mel feature and show good performances in audio scene classification task. The posterior $p(y | {\bf f}_{\bf x} ; \theta )$ is the probabilistic score of the audio scene label $y$ using the softmax:
\begin{eqnarray}
p(y | {\bf f}_{\bf x} ; \theta ) = { \exp( M_y({\bf f}_{\bf x}) ) \over \sum_{\upsilon\in {\mathcal Y}} \exp( M_{\upsilon}({\bf f}_{\bf x}) )  }
\end{eqnarray}
where $M_{v\in {\mathcal Y}}({\bf f}_{\bf x})$ is the output of ${\bf f}_{\bf x}$ obtained from the final layer of the audio scene classification network $M$.

The dataset of task1A consists of the audio scene samples recorded in a number of cities: i.e. samples of an audio scene (e.g. airport) were recorded in a number of locations (e.g. Amsterdam, London, Helsinki). In real applications, the audio scene classifier can be trained to classify $N$ audio scenes using the dataset recorded in a limited number of cities, and the classifier may be deployed to the test environments which are unseen cities in the training dataset. In such cases, the test samples can be misclassified since the classification boundary may not accurately separate the audio scene samples from unseen cities. This is illustrated in Fig. \ref{fig:cluster_example}. In this example, there are two audio scene feature points (red, blue) from three different cities (triangle, circle, rectangle). And, given an audio scene, it's highly probable that the samples from the same city are more clustered than from different cities. The black dashed line is the classification boundary trained with the audio scene data from three different cities. Here, we may have the test samples from unseen city (gray hexagon) which can be presented in the other region near the classification boundary. In this case, some of the test samples will be misclassified. In contrast to the left-side of the Fig. \ref{fig:cluster_example}, if the audio scene features are clustered independently of the cities, and the within-class distances are minimized, then we can have the classifier with more generalization ability especially when there are audio scene samples from unseen cities as shown in the right-side of the Fig. \ref{fig:cluster_example}.

\section{Proposed Algorithm}
\label{sec:proposed_algorithm}
\begin{figure}[t]
	\centering
	\centerline{\includegraphics[width=\columnwidth]{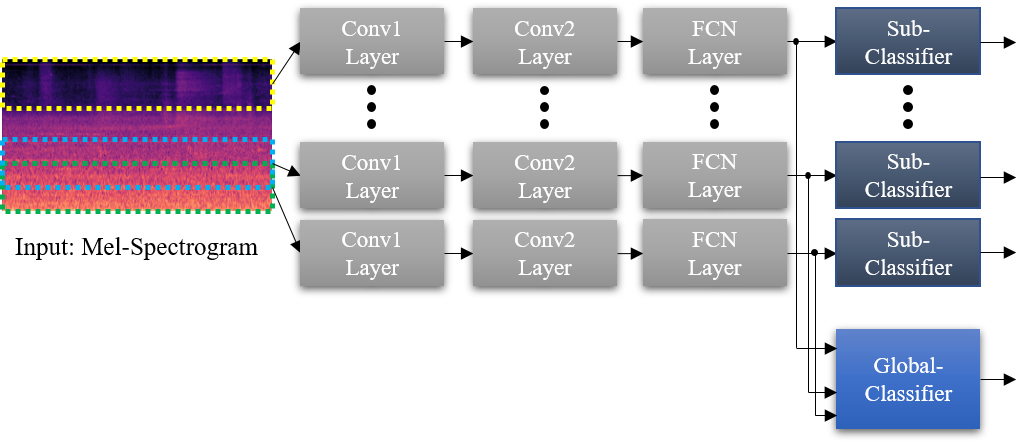}}
	\caption{Block diagram of proposed acoustic scene classification algorithm.}
	\label{fig:block_diagram}
\end{figure}

The block diagram of our proposed algorithm is illustrated in Fig. \ref{fig:block_diagram}. The overall structure is based on Sub-Spectral Net \cite{phaye2019subspectralnet} which assumes that the key patterns of each class are concentrated on the specific frequency bins, and those key patterns are effectively learned by dividing the input mel-spectrogram into the multiple sub-bands and using the CNN classifier for each sub-band independently. By dividing the input mel-spectrogram into the multiple sub-bands and learning the CNN classifier of each sub-band independently, the key patterns of specific frequency bins are effectively learned. The CNN classifiers consist of two convolution layers and one fully connected network as in the baseline network of DCASE 2019. The outputs of all FCN layers are concatenated, and they are used as the input of the global classifier.

\subsection{Factorized CNN}
Recently, the CNN-based algorithms have shown great performance and the state-of-the-art result in various areas such as image classification, audio/speech processing, speech recognition, and speaker verification \cite{torfi2018text,wan2018generalized}. As CNN-based algorithms are also a popular trend in the ASC task, most of the algorithms submitted to the DCASE 2018 are based on CNN \cite{sakashita2018acoustic,dorfer2018acoustic,zeinali2018convolutional}. The mel-spectrogram feature of audio data can be regarded as an image, and the CNN-based algorithm can be used to recognize the audio characteristics such as the phoneme and human voices.

When we train a CNN-based model for acoustic scene classification, the over-fitting problem can be occurred due to the limitation of the data even when we use simple 2-layer CNN structure is used, while learned convolution filters had a noisy pattern that was difficult to analyze. The CNN model tends to memorize all training audio scenes including the noise components which do not help classification, and it may cause the poor generalization performance.

To resolve the above-mentioned problem, we propose the factorized CNN based on low-rank matrix factorization. Low-rank matrix factorization is also widely used technique in audio signal separation \cite{sprechmann2012real,ozerov2009multichannel}, and it is based on that the mel-spectrogram of audio signal can be represented as the summation of a small number of rank 1 matrices. This leads to classify the acoustic scene with a small number of parameters for the ASC network.

\begin{figure}[t]
	\centering
	\centerline{\includegraphics[width=\columnwidth]{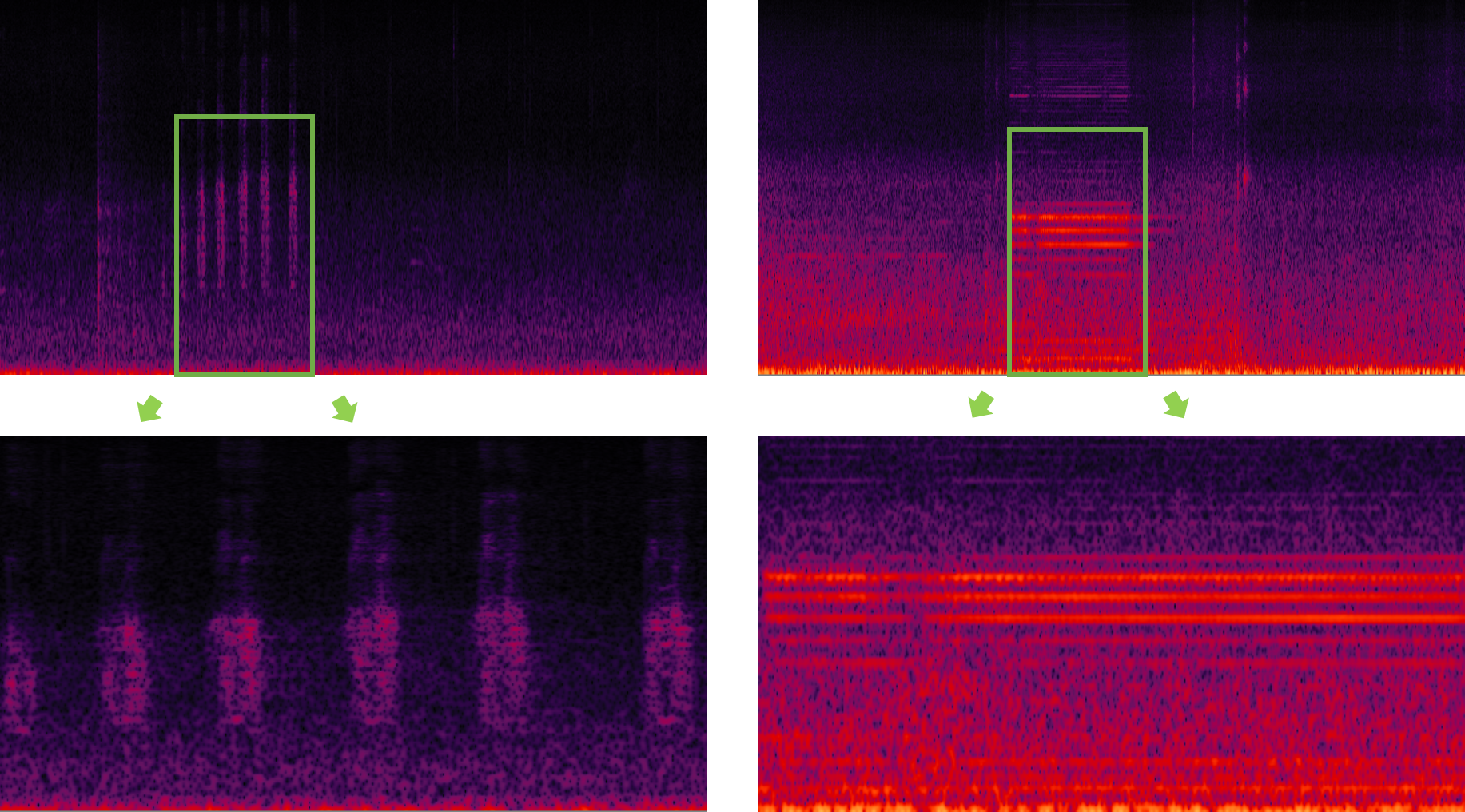}}
	\caption{The mel-spectrograms of the sound of a bird in a park and the horn of a car. (Top left) Mel-spectrogram of "park-stockholm-102-2895-a.wav" in the development dataset (Bottom left) Zoom-up of top left (Top right) Mel-spectrogram of "street\_traffic-london-271-8255-a.wav" (Bottom right) Zoom-up of top right}
	\label{fig:lowrank_example}
\end{figure}

In the classification of acoustic scenes, we can consider the following two audio elements: one is the ambient signal over a long period of time, and the other is an event signal with short period such as bird sound in a park and car horn in the road. As shown in Fig. \ref{fig:lowrank_example}, the ambient signals for each acoustic scene show faint stripes of horizontal lines in mel-spectrogram due to the statistical stationarity over time. Also, many audio examples for this kind of event signals show the pattern with the horizontal stripe, and this can be represented as a rank-1 matrix.

The detail specification of each layer is described in Figure \ref{fig:layer_spec}. Rather than using a single conv2D layer which has rectangular ($k,k$) kernel, we use two consecutive conv2D layers. These two conv2D layers have 1 dimensional kernels ($k,1$) and ($1,k$). When the rectangular ($k,k$) kernel is a rank-1 matrix and can be factorized into ($k,1$) and ($1,k$) vectors, conv2D with ($k,k$) kernel becomes equivalent to conv2D with ($k,1$) and conv2D with ($1,k$). So, this network is equivalent to conv2D layer with a rank-1 matrix kernel. With the convolution kernel of rank-1 matrix, we can reduce the over-fitting of learning noisy patterns from training data. Also, the number of model parameters can be reduced since two 1-dim kernels need $2k$ while the square kernel needs $k^2$ parameters.the square kernel needs $k^2$ parameters.

The factorized CNN originally proposed in this paper is different from the network in \cite{Wang_2017_ICCV} where it factorized the 3D convolution layer as a single intra-channel convolution and a linear channel projection.

\subsection{Large-margin loss function}
As the acoustic scene classification is the multi-class classification problem, which maximizes the probability of the correct label and minimizes that of all the others, most of the algorithms are using the cross-entropy loss function. The cross entropy between the true label $y$ and recognized output $\hat{y}$ is given as
\begin{eqnarray}
L_{\text{CE}} = - \sum_y y \log (\hat{y}).
\end{eqnarray}

\begin{figure}[t]
	\centering
	\centerline{\includegraphics[width=\columnwidth]{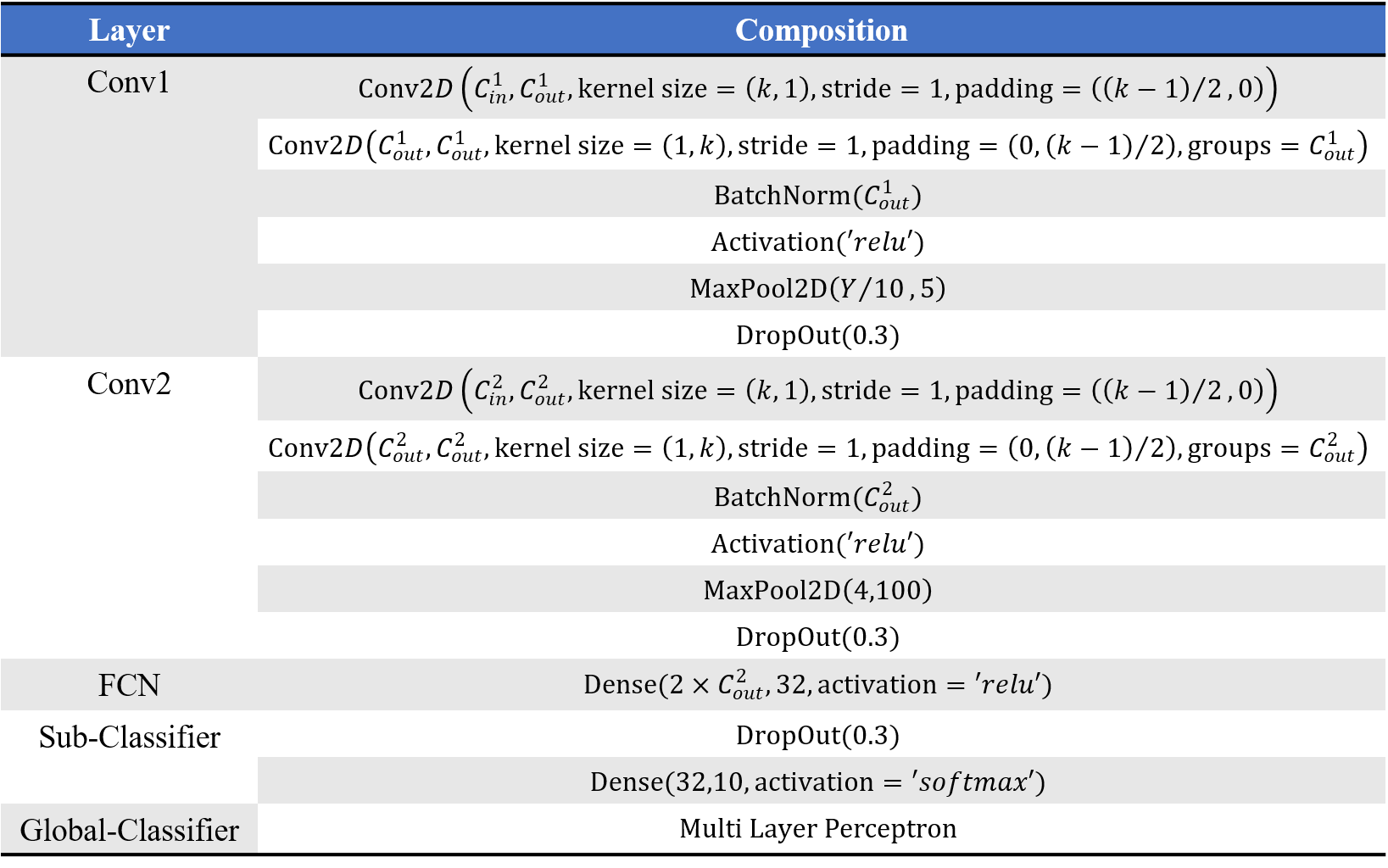}}
	\caption{The detail specifications of each block.}
	\label{fig:layer_spec}
\end{figure}

However, the cross-entropy loss only focuses on fitting or classifying the training data accurately; it does not explicitly encourage a large decision margin for classification \cite{li2019large}. Even when training the simple CNN classifier which has only 2 layers with the cross entropy loss, we can observe that the training loss converges to zero, while the test loss does not converge.

In this context, we consider a loss function similar to the triplet loss but slightly different. Triplet loss function is widely used loss function in many machine learning tasks such as person re-identification, face clustering, and speaker embedding \cite{schroff2015facenet,wan2018generalized}. It enforces the positive pairs to be closer, and the negative pairs to be further and can be expressed as
\begin{eqnarray}
L_{\text{triplet}} = \max \left(\|\mathbf{x}_a-\mathbf{x}_p\|^2-\|\mathbf{x}_a-\mathbf{x}_n\|^2+\alpha,0\right),
\end{eqnarray}
where $\mathbf{x}$ indicates the embedding vector which is the output of the final convolution layer before entering the fully connected layer, and $\mathbf{x}_a$, $\mathbf{x}_p$, and $\mathbf{x}_n$ are anchor, positive, and negative embedding vectors, respectively. $\alpha$ indicates the triplet loss margin parameter. The anchor and positive pair should come from the same class, and the anchor and negative pair should come from the different classes. In our case, we modified the sampling: we choose the positive sample from the same class but different environment (city) to cluster the samples independently of the environment as shown in Fig. \ref{fig:cluster_example}. Finally, we combine the cross-entropy and triplet losses as
\begin{eqnarray}
L_{\text{ASC}} = L_{\text{CE}} + \gamma L_{\text{triplet}}
\end{eqnarray}
where $\gamma$ is the hyper-parameter which should be tuned.

To reduce the triplet loss effectively, we should choose a distant anchor-positive pair and a close anchor-negative pair, however, it is very inefficient to figure out the distance of all pairs for choosing efficient pairs. Fortunately, the additional label which describes the city of each acoustic scene is given in DCASE 2019 dataset. The sound signals of the same acoustic scene and city is more similar than that of the same acoustic scene and different city, therefore we choose all anchor-positive pairs from the same acoustic scene and different city and all anchor-negative pairs randomly.



\subsection{Data augmentation}
Since the number of training data is limited in the development dataset, it is necessary to perform data augmentation to increase the performance of unknown data. Most of the algorithms participating in the DCASE challenge are using a deep neural network-based algorithm with high model complexity, so they are using data augmentation and it improves the performance.

We used mix-up \cite{zhang2017mixup} and spec-augment \cite{park2019specaugment} for the data augmentation. Mix-up is one of the most popular method in past DCASE 2018 challenge. It creates a new training sample by mixing a pair of two randomly chosen training samples. Spec-augment is an effective approach which shows significant performance improvement in acoustic speech recognition recently. It replaces values by zeros in randomly chosen time-frequency bands. It is also effective in acoustic scene classification task, and applied in most of the algorithms submitted in DCASE 2019 challenge.

\section{Experiment}
\label{sec:experiment}
\subsection{Dataset}
The dataset for this task is the TAU Urban Acoustic Scenes 2019 dataset, consisting of acoustic scene samples recorded in ten different European cities. For each recording location, there are 5-6 minutes of audio. The original recordings were split into segments of 10 second-log for each indifidual file. The dataset includes 10 audio scenes (e.g. 'airport', 'shopping mall'), and the development dataset contains 40 hours of data with total of 14400 segments. Here, we used 9,185 segments as a training dataset and 4,185 segments as an evaluation dataset: the split is given in the first fold of the validation set. For evaluation of unseen city, we used 1,440 segments in Milan which is not appeared in the training dataset.

\subsection{Setup}
Our ASC framework was implemented as python script using Torch library, and all experiments were conducted on a GeForce GTX TITAN X GPU with 12Gb RAM. Given the 10 second-long stereo audio input  sampled as 48kHz, we extracted 40 and 200 logmel features with stereo channel for the input of CNN. And, as in \cite{phaye2019subspectralnet}, we set the sub-spectrogram size and overlap length to 20 and 10, respectively. The input/output channel sizes of CNN structure were set as follows: $C_{in}^1=2, C_{out}^1=64, C_{in}^2=64, C_{out}^2=64$ for 40 logmel case and $C_{in}^1=2, C_{out}^1=32, C_{in}^2=32, C_{out}^2=64$ for 200 logmel case. The kernel size $k$ was set to $7$ for all logmel cases. All of the convolution filters and weight matrices in dense layers were initialized by kaiming normal and xavier normal functions in pyTorch, respectively. The learning rate was set to $0.001$ with Adam optimizer. The hyper-parameters of triplet loss margin and the balance coefficient between the cross-entropy and triplet losses were respectively set as $\alpha=0.2$, and $\gamma=10$.

\subsection{Result}
\begin{table}[t]
	\centering
	\begin{tabular}{|l|c|c|c|c|}
		\hline\hline
		\multicolumn{1}{|c|}{Validation dataset} & \multicolumn{2}{c|}{Overall}    & \multicolumn{2}{c|}{Unseen city}     \\ \hline
		\multicolumn{1}{|c|}{Feature (Logmel number)}            & 40      & 200     & 40      & 200     \\ \hline\hline
		SubSpectralNet \cite{phaye2019subspectralnet}                          & 68.93          & 73.44          & 58.29          & 68.71          \\ \hline
		FCNN                                     & 71.15          & 75.97          & 59.89          & 70.32          \\ \hline
		FCNN-mixup                               & 71.57          & 76.25          & 62.19          & 68.70          \\ \hline
		FCNN-spec                                & 71.85          & 76.44          & 63.90          & 71.74          \\ \hline
		FCNN-mixup-spec                          & 72.76          & 75.97          & 62.62          & 70.40          \\ \hline
		FCNN-triplet                             & 72.67          & 76.61          & 63.07          & 70.24          \\ \hline
		FCNN-triplet-spec                        & \textbf{73.14} & \textbf{77.19} & \textbf{64.23} & \textbf{72.38} \\ \hline
		DCASE 2019 Rank 1 \cite{Chen2019}		& \multicolumn{2}{c|}{85.10} 	& \multicolumn{2}{c|}{-}				\\ \hline\hline
	\end{tabular}
	\caption{Classification accuracies (\%) of the proposed algorithm with different settings and the baseline CNN algorithm.}
	\label{performance_table}
\end{table}
The DCASE 2019 development dataset includes recordings from ten cities while the training subset contains only nine cities: the recordings of Milan are not included. By checking the performance of unseen city, we can measure the generalization ability of the model. Thus, we measured not only the accuracy of overall validation dataset but also the accuracy of the dataset of unseen city.

Experimental results are summarized in Table \ref{performance_table}. Since the classification accuracy of the evaluation set was saturated before the 100-th epoch, we trained all models for 200 epochs and averaged over the accuracies of last 10 epochs. For each evaluation of the algorithms, the model training was conducted twice, and the accuracies were also averaged.

For the overall dataset, the factorized CNN (FCNN) improved the performance over the baseline network by 2.22\% and 2.53\% for 40 and 200 logmel cases, respectively. Further improvement was observed when the mix-up and spec-augment were used. By adopting the proposed loss function combined with the triplet loss, the FCNN performance was improved by 1.52\% and 0.64\%. When the spec-augment was applied to the FCNN with the proposed loss function, we obtained the best performance improvement from the baseline by 4.21\% and 3.75\%. In the experiments, the mix-up couldn't be used with triplet loss since the labels of augmented data using mix-up are not discrete. For the evaluation of only unseen city, more performance improvement was achieved compared to the evaluation of entire dataset. The total performance improvements for logmel 40 and 200 cases are respectively 5.94\% and 3.67\%. For 40 logmel case, the relative performance improvement is 41\% which is much bigger than that in evaluating the entire dataset. For the comparisons, we added the DCASE 2019 challenge top rank result in Table \ref{performance_table}. Note that the performance comparison with DCASE 2019 Challenge Rank 1 system \cite{Chen2019} is unfair since the system characteristics such as the number of input features, model size, usage of ensemble and so on are different from ours, but we enumerated it as a reference. 

\begin{table}[]
	\centering
	\begin{tabular}{|c|c|c|}
		\hline\hline
		Input feature               & Network        & \# of param \\ \hline\hline
		\multirow{3}{*}{Logmel 40}  & DCASE baseline & 117K        \\ \cline{2-3} 
		& SubSpectralNet \cite{phaye2019subspectralnet} & 331K        \\ \cline{2-3} 
		& Proposed FCNN  & 113K        \\ \hline
		\multirow{2}{*}{Logmel 200} & SubSpectralNet \cite{phaye2019subspectralnet}& 2,541K      \\ \cline{2-3} 
		& Proposed FCNN  & 871K        \\ \hline
		\multicolumn{2}{|c|}{DCASE 2019 Rank 1 \cite{Chen2019}} & 48M\\ \hline\hline
	\end{tabular}
	\caption{The number of parameters for the DCASE baseline, SubSpectralNet, and our proposed FCNN}
	\label{model_parameters}
\end{table}

Since the number of parameters of CNN is one of the main criteria, we briefly compared the number of model parameters in Table \ref{model_parameters}. As in Table \ref{model_parameters}, the number of proposed FCNN parameters is about one-third of SubSpectralNet, and it is almost the same as DCASE baseline network. 

\section{Conclusion}
\label{sec:conclusion}
In this paper, an acoustic scene classification algorithm based on a large-margin factorized CNN is proposed. The motivation of a factorized CNN is that the key patterns in the mel-spectrogram including the long-term ambient and short-term event sounds are low-rank, and thus the factorized CNN can effectively learn these key patterns with a reduced number of model parameters. To increase the generalization performance of the learned model, a modified triplet loss is combined with the cross-entropy loss function. Experimental results show that our proposed algorithm outperforms a conventional simple CNN-based algorithm with smaller model complexity. Further improvement on unseen data is also observed, and it can be considered that the proposed algorithm has better generalization performance.


%
%
%

\bibliographystyle{IEEEtran}
\bibliography{refs}
%
%
%
%
%
%
%
%
%

\end{sloppy}
\end{document}